\documentstyle[psfig]{mn}

\newif\ifAMStwofonts

\newcommand{\msun}{{\mathrm M_\odot}}
\newcommand{\pop}{{\mbox{Pop III}}}
\newcommand{\lya}{{{\mathrm Ly}\alpha}}
\newcommand{\dd}{{\mathrm {d}}}

\newcommand{\mcrit}{{M_{\mathrm crit}(z)}}

\newcommand{\tcrit}{{T_{\mathrm crit}}}
\newcommand{\psion}{{\psi_{\mathrm on}}}

\newcommand{\hz}{{\mathrm {Hz}}}
\newcommand{\yr}{{\mathrm{yr}}}
\newcommand{\cm}{{\mathrm{cm}}}
\newcommand{\ev}{{\mathrm{eV}}}
\newcommand{\cmt}{{\mathrm{cm}^{-3}}}

\newcommand{\mpc}{{\mathrm{Mpc}}}
\newcommand{\kms}{{\mathrm{km~s^{-1}}}}
\newcommand{\se}{{\mathrm{s}}}
\newcommand{\kelvin}{{\mathrm{K}}}
\newcommand{\ang}{{\mathrm{\AA}}}
\newcommand{\hi}{{\mbox{H\,{\sc i}}}}
\newcommand{\hii}{{\mbox{H\,{\sc ii}}}}
\newcommand{\hei}{{\mbox{He\,{\sc i}}}}
\newcommand{\heii}{{\mbox{He\,{\sc ii}}}}
\newcommand{\heiii}{{\mbox{He\,{\sc iii}}}}

\ifoldfss
  \ifCUPmtlplainloaded \else
    \NewTextAlphabet{textbfit} {cmbxti10} {}
    \NewTextAlphabet{textbfss} {cmssbx10} {}
    \NewMathAlphabet{mathbfit} {cmbxti10} {} 
    \NewMathAlphabet{mathbfss} {cmssbx10} {} 
  \fi
  \ifAMStwofonts
    \ifCUPmtlplainloaded \else
      \NewSymbolFont{upmath} {eurm10}
      \NewSymbolFont{AMSa} {msam10}
      \NewMathSymbol{\upi}     {0}{upmath}{19}
      \NewMathSymbol{\umu}     {0}{upmath}{16}
      \NewMathSymbol{\upartial}{0}{upmath}{40}
      \NewMathSymbol{\leqslant}{3}{AMSa}{36}
      \NewMathSymbol{\geqslant}{3}{AMSa}{3E}

      \let\leq=\leqslant 
      \let\geq=\geqslant \let\ge=\geqslant
    \fi
  \fi
\fi 

\ifnfssone
  \newmathalphabet{\mathit}
  \addtoversion{normal}{\mathit}{cmr}{m}{it}
  \addtoversion{bold}{\mathit}{cmr}{bx}{it}
  \newmathalphabet{\mathbfit} 
  \addtoversion{normal}{\mathbfit}{cmr}{bx}{it}
  \addtoversion{bold}{\mathbfit}{cmr}{bx}{it}
  \newmathalphabet{\mathbfss} 
  \addtoversion{normal}{\mathbfss}{cmss}{bx}{n}
  \addtoversion{bold}{\mathbfss}{cmss}{bx}{n}
  \ifAMStwofonts
    \ifCUPmtlplainloaded \else
      \UseAMStwoboldmath
      \makeatletter
      \new@mathgroup\upmath@group
      \define@mathgroup\mv@normal\upmath@group{eur}{m}{n}
      \define@mathgroup\mv@bold\upmath@group{eur}{b}{n}
      \edef\UPM{\hexnumber\upmath@group}
      \new@mathgroup\amsa@group
      \define@mathgroup\mv@normal\amsa@group{msa}{m}{n}
      \define@mathgroup\mv@bold\amsa@group{msa}{m}{n}
      \edef\AMSa{\hexnumber\amsa@group}
      \makeatother
      \mathchardef\upi="0\UPM19
      \mathchardef\umu="0\UPM16
      \mathchardef\upartial="0\UPM40
      \mathchardef\leqslant="3\AMSa36
      \mathchardef\geqslant="3\AMSa3E

      \let\leq=\leqslant 
      \let\geq=\geqslant \let\ge=\geqslant
    \fi
  \fi
\fi 

\ifnfsstwo
  \DeclareMathAlphabet{\mathbfit}{OT1}{cmr}{bx}{it}
  \SetMathAlphabet\mathbfit{bold}{OT1}{cmr}{bx}{it}
  \DeclareMathAlphabet{\mathbfss}{OT1}{cmss}{bx}{n}
  \SetMathAlphabet\mathbfss{bold}{OT1}{cmss}{bx}{n}
  \ifAMStwofonts
    \ifCUPmtlplainloaded \else
      \DeclareSymbolFont{UPM}{U}{eur}{m}{n}
      \SetSymbolFont{UPM}{bold}{U}{eur}{b}{n}
      \DeclareSymbolFont{AMSa}{U}{msa}{m}{n}
      \DeclareMathSymbol{\upi}{0}{UPM}{"19}
      \DeclareMathSymbol{\umu}{0}{UPM}{"16}
      \DeclareMathSymbol{\upartial}{0}{UPM}{"40}
      \DeclareMathSymbol{\leqslant}{3}{AMSa}{"36}
      \DeclareMathSymbol{\geqslant}{3}{AMSa}{"3E}

      \let\leq=\leqslant 
      \let\geq=\geqslant \let\ge=\geqslant
    \fi
  \fi
\fi 

\ifCUPmtlplainloaded \else
  \ifAMStwofonts \else 
    \def\upi{\pi}
    \def\umu{\mu}
    \def\upartial{\partial}
  \fi
\fi

\title{The Contribution of the First Stars to the Cosmic Infrared Background}
\author[M. R. Santos {\rm et al.}]
       {Michael R. Santos,$^{1}$ Volker Bromm$^2$ and Marc
       Kamionkowski$^1$ \\
$^1$California Institute of Technology, Mail Code 130-33, 
Pasadena, CA 91125, U.S.A.\\
$^2$Harvard-Smithsonian Center for Astrophysics,
60 Garden Street, Cambridge, MA 02138, U.S.A.}

\pagerange{\pageref{firstpage}--\pageref{lastpage}}
\pubyear{2001}

\begin{document}

\maketitle

\label{firstpage}

\begin{abstract}
We calculate the contribution to the cosmic infrared background from
very massive metal-free stars at high redshift.  We explore two
plausible star-formation models and two limiting cases for the
reprocessing of the ionizing stellar emission.  We find that
\mbox{Population III} stars may contribute significantly to the cosmic
near-infrared background if the following conditions are met: (i) The
first stars were massive, with $M\ga 100~\msun$.  (ii) Molecular
hydrogen can cool baryons in low-mass haloes. (iii) Pop III star
formation is ongoing, and not shut off through negative feedback
effects.  (iv) Virialized haloes form stars at about 40 per cent
efficiency up to the redshift of reionization, $z\sim7$.  (v) The
escape fraction of the ionizing radiation into the intergalactic
medium is small.  (vi) Nearly all of the stars end up in massive black
holes without contributing to the metal enrichment of the Universe.
\end{abstract}

\begin{keywords}
cosmology: theory -- early Universe -- galaxies: formation -- 
stars: formation -- diffuse radiation -- infrared: general.
\end{keywords}

\section{Introduction}

It has long been realized that observations of the cosmic infrared
background (CIRB) can place important constraints on the energy
production associated with the formation of cosmological structure
(see Hauser \& Dwek 2001 and references therein).  The cosmic
background is the locally measured radiation density from all
extragalactic sources.  Starlight dominates the CIRB in the near-IR,
whereas the mid- and far-IR backgrounds result primarily from dust
emission (e.g., Dwek et al. 1998).

Observational efforts to measure the near-IR CIRB are hampered by
strong and uncertain foreground contamination from zodiacal dust.
Cambr\'{e}sy et al. (2001), using data from the Diffuse Infrared
Background Experiment (DIRBE) on the {\it COBE} satellite, found that
the integrated light from all galaxies (estimated from deep surveys)
is not sufficient to account for the measured CIRB at
$1.25~\umu\mathrm{m}$ ($J$ band) or $2.2~\umu\mathrm{m}$ ($K$ band).
Wright \& Johnson (2001) analyzed the same DIRBE data as Cambr\'{e}sy
et al. (2001), but subtracted a different zodiacal light model; they
find a $J$-band CIRB that is consistent with the contribution from
galaxies, but their $K$-band CIRB value is larger than the background
inferred from galaxy counts.  Totani et al. (2001) model the
contribution from galaxies missed by deep galaxy surveys and find that
it is unlikely to be greater than 30 per cent of the measured light.

Bond, Carr \& Arnett \shortcite{bon86} suggested that the first
generation of stars in the Universe, called \mbox{Population III}
(\pop) stars because they are assumed to have near-zero metallicity,
may contribute significantly to the cosmic background.  The
cosmological impact of \pop\ stellar radiation has been addressed
before (e.g., Carr, Bond \& Arnett 1984; Bond et al. 1986; Couchman \&
Rees 1986; Haiman \& Loeb 1997; Ciardi et al. 2001).  Recent
theoretical advances have improved our understanding of the physics of
\pop\ star formation (see Barkana \& Loeb 2001 and references
therein); in particular, three-dimensional numerical simulations
suggest that the primordial initial mass function (IMF) may have been
concentrated at stellar \mbox{masses $\ga 100~\msun$} (Bromm, Coppi \&
Larson 1999, 2002; Abel, Bryan \& Norman 2000).  Stars with
\mbox{masses $\ga 100~\msun$}, which we will call `very massive'
\cite{car84}, have spectra and luminosities that asymptotically
approach the blackbody form and the Eddington limit, respectively
(e.g., Bromm, Kudritzki \& Loeb 2001b).

In this paper, we investigate whether a significant fraction of the
near-IR CIRB may come from very massive \pop\ stars.  Since these
stars are very luminous per unit stellar mass over a short lifetime,
they may contribute substantially to the CIRB.  We use two different
models for the formation of dark matter (DM) haloes to calculate the
rate at which baryons are processed through haloes with virial
temperatures high enough to allow baryonic cooling
(\S\ref{sfrsec}). In addition, we consider two possibilities for the
re-processing of the stellar emission by the gas in the halo and by
the intergalactic medium (\S\ref{specsec}).  In \S\ref{cirbsec} we
show our model results and the observational data.  Section
\ref{discsec} contains a discussion of our results and compares them
with the data.

We adopt the following values for the cosmological parameters:
$h=0.7$, $\Omega_{m}=0.3$, $\Omega_\Lambda=0.7$, $\Omega_{\mathrm B}
h^2=0.019$, $\sigma_8=0.9$.  Here $h$ is the dimensionless Hubble
constant, $H_0=100h~\kms\mpc^{-1}$; $\Omega_{m}$, $\Omega_\Lambda$ and
$\Omega_{\mathrm B}$ are the total matter, cosmological constant and
baryon density in units of the critical density and $\sigma_8$ gives
the normalization of the power spectrum on the $8h^{-1}$Mpc scale.

\vspace{-0.25 cm}

\section{Star-formation rate} \label{sfrsec}

\subsection{Overview of the star-formation model} \label{sfroversec}

We assume that the star-formation rate (SFR) of \pop\ stars is related
to the cooling of baryons in collapsed dark-matter haloes (henceforth
just `haloes').  The cooling rate of baryons is a very strong function
of baryon temperature near certain temperature thresholds: above these
thresholds a new cooling mechanism can act within the gas.  For a
given choice of cooling mechanism, we would like to calculate the rate
at which baryons are heated above the associated temperature
threshold, and thus may collapse to high density and form stars.

If baryons in a halo are shock-heated to the virial temperature of
their halo, the temperature of those baryons is related to the mass of
that halo according to $T\propto M^{2/3}$.  So as a proxy for the
temperature history of the baryons, we may calculate the mass history
of haloes, assuming a constant ratio of baryonic matter to dark matter
in all haloes, and that baryons below the cooling threshold are always
heated to the virial temperature of their halo.  In this picture
haloes above a critical mass may form stars, because those haloes have
heated their gas above the critical temperature, and haloes below the
critical mass do not form any stars.  Additionally, we parametrize the
fraction of eligible baryons that actually do form stars with a
constant star formation efficiency, $\eta$.

The mass-assembly history of dark-matter haloes is computed using the
extended Press-Schechter formalism \cite{lac93}, where the
$\sigma(M)$--$M$ relation is evaluated with
the power spectrum of Eisenstein \& Hu (1999).
Here $\sigma(M)$ is the standard deviation of the linear density
field smoothed on scales containing a mean mass $M$. Our choices of
$\mcrit$ are described in \S \ref{critsec}.  

We consider two models for the SFR.  In both models, star formation is
triggered in a halo as it accumulates enough mass to put it above the
threshold mass $\mcrit$, but the models differ in how they treat
additional material that merges into such haloes.  In the `ongoing'
model, all gas in haloes with $M\geq\mcrit$ is eligible for star
formation.  Conversely, the `single-burst' model only allows star
formation in haloes that have not previously formed stars and have no
progenitor halo that has formed stars, i.e., both merging haloes are
crossing the critical threshold for the first time.

\subsection{Specific star-formation models}

\subsubsection{Ongoing star formation} \label{sfronsec}

In the ongoing model, the SFR in a single halo is proportional to the
growth (in mass) of that halo, after it has become more massive than
$\mcrit$.  Star formation does not occur in a quiesciently evolving
(i.e., non-accreting) halo; the star formation is not {\it
continuous} in the usual sense.  But star formation is not inhibited
after the first generation of \pop\ stars is born in a halo, which
assumes that radiation and mechanical outflows from the life and death
of the star(s) have no impact (zero feedback) on future star
formation.

The SFR in the ongoing model is proportional to the time derivative of
the total mass contained in haloes that have mass $M\geq\mcrit$.  The
SFR per comoving volume, $\psion$, is
\begin{equation}
\psion(z,\mcrit) = 
\eta \frac{\Omega_{\mathrm B}}{\Omega_{m}} 
\frac{{\mathrm {d}}}{{\mathrm {d}} t} 
\int_{\mcrit}^{\infty} {{\mathrm {d}}} M~M 
\frac{{\mathrm{d}} n_{\mathrm PS}}{{\mathrm{d}} M}(M,z).
\end{equation}
Here $n_{\mathrm PS}\equiv n_{\mathrm PS}(M,z)$ is the comoving number
density of haloes of mass $M$ at redshift $z$, given by Press \&
Schechter \shortcite{pre74}.  The integral expresses the collapsed
mass per comoving volume contained in haloes above the critical mass.
The time derivative converts this to a mass rate, and the prefactors
convert from total mass to stellar mass.  

\subsubsection{Single-burst star formation} \label{sfrsingsec}

In the single-burst model, a halo forms stars when the accretion of
matter pushes its mass above the critical mass, $\mcrit$; that is the
only time that the halo will form stars, regardless of additional
mergers and mass accretion.  This is a very extreme model which
assumes \pop\ stars make a long-term change in their surroundings
which permanently prohibits future \pop\ stars from forming there.
This extreme model could be realistic if \pop\ stars heavily enrich
the interstellar medium with metals.  Then, though stars will likely
eventually form out of that material, \pop\ stars will not.  The SFR
may lie somewhere between the prediction of this model, and the
prediction of the ongoing model discussed above (in which all baryons
in haloes with $M>\mcrit$ are eligible to form stars).

The SFR in the single-burst model can be calculated with the extended
Press-Schechter formalism: A halo grows in mass through discrete
mergers and gradual accretion, which is treated as a series of small
mergers in extended Press-Schechter theory.  As an example, take a
halo with mass $M_1$ at some redshift $z$ that merges with mass
$\Delta M$ for a total mass of $M_2=M_1+\Delta M$ at redshift
$z+{\mathrm d}z$, where ${\mathrm d}z$ is small and negative, i.e., a
short time later.  If $M_1<\mcrit$ and $M_2\geq\mcrit$, then, provided
$\Delta M<\mcrit$, the halo undergoes a burst of star formation.  We
parametrize the mass of stars formed, $M_*$, by
\begin{equation}
M_* = \eta \frac{\Omega_{\mathrm B}}{\Omega_{m}} M_2.
\end{equation}
The SFR per comoving volume, $\psi_{\mathrm burst}$, is
\begin{displaymath}
\psi_{\mathrm burst}(z,\mcrit)=\frac{1}{2}
\frac{\eta~\Omega_{\mathrm B}}{\Omega_{m}}
\end{displaymath}
\begin{equation}
\times
\int_{0}^{\mcrit}{\mathrm{d}} M_1\frac{{\mathrm{d}} n_1}{{\mathrm{d}} M_1}
\int_{\mcrit}^{\mcrit+M_1}{\mathrm{d}} M_2~M_2
\frac{{\mathrm{d}}^2P}{{\mathrm{d}} M_2~{\mathrm{d}} t},
\end{equation}
where $n_1\equiv n_{\mathrm PS}(M_1,z)$ and $P\equiv P(M_1,M_2,z)$ is
the probability that a halo with mass $M_1$ merges to a new mass
$M_2>M_1$ at redshift $z$ \cite{lac93}.  The prefactor of $1/2$
corrects for double counting in the integrals (for fixed $M_1$ and
$\Delta M$, the integral as written counts both $M_1+\Delta M=M_2$ and
$\Delta M+M_1=M_2$). \footnote{There is some ambiguity in the SFR
arising from the extended Press-Schechter merger rates.  We defer
discussion of that to future work, and here evaluate the SFR as
written.}
A similar approach to model the history of star formation at high
redshifts has been discussed by Barkana \& Loeb (2000).

\subsection{Critical halo mass for star formation} \label{critsec}

There is a relationship between the mass of a halo and its virial
temperature.  Using the assumptions we made in \S\ref{sfroversec}, we
can substitute a critical temperature for baryonic cooling, $\tcrit$,
into that relation to find the corresponding critical mass:
\begin{displaymath}
\mcrit=0.94\times 10^{8}~\msun\left(\frac{h}{0.7}\right)^{-1}
\left(\frac{\Omega_{m}}{0.3}\right)^{-1/2}
\end{displaymath}
\begin{equation}
\hspace{2 cm}\times
\left(\frac{1+z}{10}\right)^{-3/2}
\left(\frac{\mu}{0.6}\right)^{-3/2}
\left(\frac{\tcrit}{10^4~\kelvin}\right)^{3/2},
\end{equation}
where $\mu$ is the mean molecular weight ($\mu=0.6$ for
ionized gas [$\tcrit\ga10^4~\kelvin$] and $\mu=1.2$ for neutral gas
[$\tcrit\la10^4~\kelvin$]) \cite{bar01}.

We seek to model only two cooling processes: radiative cooling from
molecular hydrogen and radiative cooling from atomic hydrogen.  For
molecular cooling, we choose $\tcrit=400~\kelvin$ (e.g., Tegmark et
al. 1997; Abel et al. 1998).  For atomic cooling, we choose
$\tcrit=10^4~\kelvin$ \cite{bar01}.

\subsection{Limits on the abundance of \pop\ stars} \label{omegasec}

A realistic model of \pop\ star formation must conform to some
observational constraints.  The matter density processed through \pop\
stars as a fraction of the critical density is
\begin{equation}
\Omega_{\mathrm{III}}(z) = \frac{1}{\rho_{\mathrm c}}~
\int_{z}^{\infty}\psi(z')\left|\frac{\dd t'}{\dd z'}\right| \dd z',
\end{equation}
with
\begin{displaymath}
\left|\frac{\dd t}{\dd z}\right| = \left[H_0 (1+z) E(z)\right]^{-1},
\end{displaymath}
and, in a flat universe,
\begin{displaymath}
E(z)=[\Omega_{m}(1+z)^3+\Omega_\Lambda
]^{1/2}.
\end{displaymath}
Here,
$\psi$ is the SFR and $\rho_{\mathrm c}$ is the density required to
close the Universe.  A strong upper limit on $\Omega_{\mathrm{III}}$
is that it must not exceed $\Omega_{\mathrm B}$.  Below we discuss two
constraints on the possible products of \pop\ stars, black holes and
heavy elements (see also Schneider et al. 2002). 

\subsubsection{Black holes} \label{bhsec}

A nonrotating star with mass above $\sim 260~\msun$ is predicted to
evolve directly to a black hole without any metal ejection
\cite{fry01}.  The fate of rotating stars in that mass range is
unknown, but may be similar.  If so, and if the \pop\ IMF were
concentrated at masses above $\sim 260~\msun$, most of the baryons
once in \pop\ stars would now be in black holes.  Such black holes
would populate the halo of our galaxy, but they would be very
difficult to detect unless they have a companion; e.g., the MACHO
microlensing study places no constraint on the fraction of the Milky
Way halo contained in $\ga 100~\msun$ black holes \cite{alc01}.
Detailed theoretical studies may eventually reveal the fraction of
such black holes that currently have a companion object, and searches
for those binary systems, or future microlensing studies, could lead
to constraints on $\Omega_{\mathrm BH}$ (e.g., Agol \& Kamionkowski
2001).  But currently even if $\Omega_{\mathrm BH}=\Omega_{\mathrm
III}$ there is no additional constraint on $\Omega_{\mathrm III}$ from
limits on black holes.

\subsubsection{Metal enrichment}

A nonrotating star with mass between $\sim 140~\msun$ and $\sim
260~\msun$ is expected to end its evolution in a pair-instability
supernova, which completely disrupts the star and leaves no remnant
\cite{heg01}.  In this case all of the nucleosynthetic products of the
star are ejected, and potentially pollute the intergalactic medium
(IGM) (cf. \S\ref{bhsec}, where all were swallowed into a black
hole). Thus metal abundances in the Universe, especially at high
redshift, are limits on the number of \pop\ stars which resulted in
supernovae, given assumptions about the mixing of the metals.

If a fixed fraction $\epsilon_{\mathrm pi}$ (`pi' for `pair
instability') of the mass converted into \pop\ stars is expelled in
pair-instability supernovae, the nucleosynthetic yield calculation from
Heger \& Woosley (2002) may be used to determine the
contribution from \pop\ stars to the metallicity of the IGM.  For
\begin{equation}
\left[i\right] \equiv \log_{10}\left(\frac{f^U_i}{f^\odot_i}\right),
\end{equation}
where $i$ is a species and $f^U_i$ and $f^\odot_i$ are the mass
fractions of species $i$ in the Universe and Sun, respectively.
Similarly, if we denote the ejected mass fraction of species $i$ 
by $f^E_i$, then
\begin{equation}
f^U_i = \epsilon_{\mathrm pi}
\frac{\Omega_{\mathrm III}}{\Omega_{B}}f^E_i
\end{equation}
and
\begin{equation}
\left[i\right] = \log_{10}\left(\epsilon_{\mathrm pi}
\frac{\Omega_{\mathrm III}}{\Omega_{\mathrm B}}P_i\right),
\end{equation}
where $P_i\equiv f^E_i/f^\odot_i$ is the production factor of species
$i$.  Heger \& Woosley (2002, table 4) calculated $P_i$ for many
isotopes and stellar masses. We here take
$\sim 200~\msun$ as a fiducial mass for a
pair-instability supernova progenitor.

A few values of interest from their table are $P_i(i=
\mbox{${}^{12}\mathrm{C}$},\mbox{${}^{16}\mathrm{O}$},
\mbox{${}^{24}\mathrm{Mg}$},\mbox{${}^{28}\mathrm{Si}$},
\mbox{${}^{56}\mathrm{Fe}$})
=(13.2,45.8,85.7,353,49.8)$.  In general, the highest values of $P_i$ are
for the alpha elements.

Observations of the abundance of species $i$ in the high-redshift
universe are sometimes quoted in slightly different notation from [i].
Assuming $\epsilon_{\mathrm
pi}\Omega_{\mathrm{III}}\ll\Omega_{\mathrm{B}}$,
$[i/\mathrm{H}]\simeq[i]$.  Using tabulated values for $f^E_i$ (Heger
\& Woosley 2002, table 3), it is straightforward to calculate
$\Omega_{i}=f^U_i \Omega_{\mathrm B}$.  With assumptions about the
mixing of the ejecta from \pop\ stars, measurements of metals in the
IGM at high redshifts (or the metallicities of stars formed at high
redshifts) can place a limit on a combination of the IMF of \pop\
stars (which we simplify into the parameter $\epsilon_{\mathrm pi}$)
and $\Omega_{\mathrm III}$.  For example, if $\epsilon_{\mathrm pi}=1$
and $f_{\mathrm B}\equiv\Omega_{\mathrm III}/\Omega_{\mathrm
B}\simeq3\times10^{-3}$, then \pop\ stars would, in the mean, enrich
the Universe to the solar abundance of silicon already at a redshift $\ga 7$.
In contrast, if
$\epsilon_{\mathrm pi}=0$, then metallicity measurements place no
constraints on the abundance of \pop\ stars.

\subsection{Star-formation model results} \label{sfrmodsec}

\begin{figure}
\psfig{file=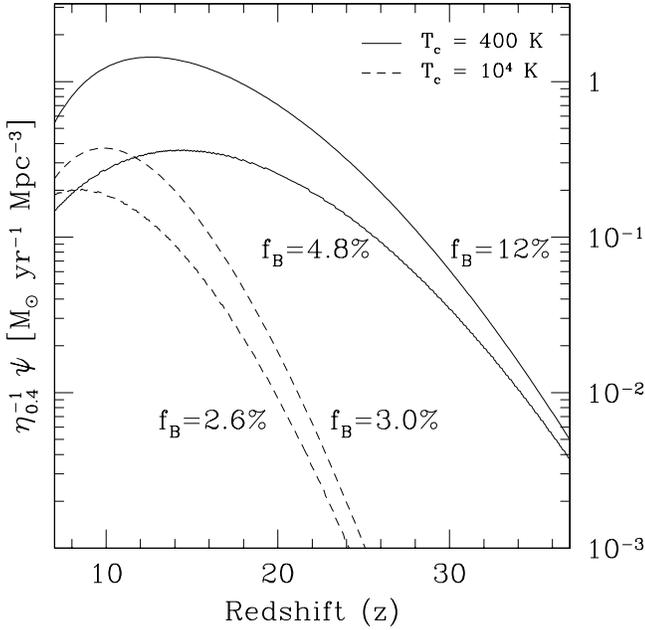,width=3.4in}
\caption{The star-formation rate (SFR) for the ongoing and
single-burst models, for critical temperatures of $T_c=400$ ({\it
solid lines}) and $10^4~\kelvin$ ({\it dashed lines}).  For each $T_c$
the upper curve is the ongoing SFR and the lower curve is the
single-burst SFR.  All curves are for $\eta=0.4$; each curve is
labelled by the value of $f_{\mathrm B}\equiv\Omega_{\mathrm III}(z=7)
/ \Omega_{\mathrm B}$ it implies.}
\label{sfretafig}
\end{figure}

\begin{figure}
\psfig{file=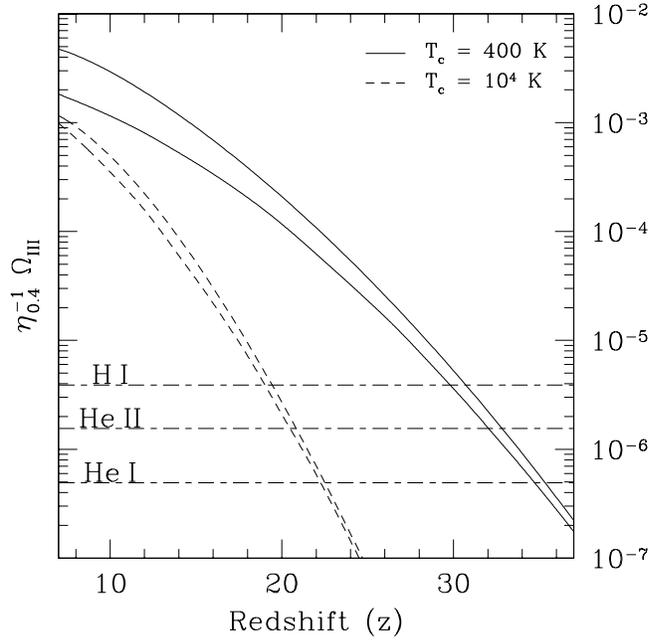,width=3.4in}
\caption{The matter density processed through \pop\ stars in units of
the critical density, as a function of redshift.  The four curves
correspond to the four different SFRs of Fig. \ref{sfretafig}.  The
horizontal dashed lines labelled \hi, \hei\ and \heii\ correspond to
the $\Omega_{\mathrm III}$ required for \pop\ stars to produce 10
ionizing photons per particle of each species.  These lines are only
relevant for the $f_{\mathrm esc}=1$ case; see section \ref{specsec}.}
\label{omegafig}
\end{figure}

To compare our models, we present the SFR as a function of redshift in
\mbox{Fig. \ref{sfretafig}}.  We assume that \pop\ star formation
shuts off at $z_{\mathrm end}\simeq7$; this may be due to
photoevaporation of low-mass haloes due to reionization \cite{bar99}
or some other process.  \mbox{Figure \ref{sfretafig}} shows the SFRs
normalized to $\eta=0.4$.  The curves are labelled with the
corresponding value of $f_{\mathrm B}$.

\mbox{Figure \ref{omegafig}} shows $\Omega_{\mathrm III}(z)$ for the
SFRs of \mbox{Figure \ref{sfretafig}}.  The horizontal lines are
explained in \S\ref{discsec}.

\section{Spectra} \label{specsec}

\subsection{Input stellar spectrum}

Metal-free very massive stars have spectra similar to a $\sim
10^5~\kelvin$ blackbody spectrum \cite{bro01c}.  Most of the energy is
then radiated in photons with energies $>13.6~\ev$.  There are also
many photons produced capable of ionizing \heii\ (but see Schaerer
2002).  The specific luminosity per solar mass of very massive stars
is almost independent of stellar mass \cite{bro01c}. We will therefore
take the spectrum of a $1000~\msun$ star as our fiducial input \pop\
stellar spectrum.

When a \pop\ star becomes luminous, we expect the host halo to
contain gas not incorporated into stars; we call this gas the
`nebula.'  We refer to the gas outside the collapsed halo as the IGM.
The physical environment of \pop\ star formation will likely undergo a
transition from dense gas in the nebula to a diffuse IGM, but we will
model all gas as either part of the nebula or part of the IGM.  {\it
We also assume that both the nebula and IGM are entirely free of
dust.}

Since the nebula and IGM are neutral in the absence of stellar
radiation, both have the potential to play an important role in
reprocessing ionizing photons from a star.  We examine two limiting
cases for the importance of each of these phases: In the first case,
\pop\ stars are enshrouded in dense nebulae, and all of the
reprocessing of ionizing radiation takes place in the halo (the IGM
still plays a role in scattering $\lya$ photons).  In this case, the
escape fraction of ionizing photons from the nebula, $f_{esc}$, is
zero.  In the second case, the nebula plays no role and $f_{esc}=1$;
all reprocessing occurs in the IGM.  In the rest of this section, we
discuss the reprocessed spectrum of \pop\ stellar radiation for each
of these cases.

\subsection{No escape of ionizing radiation into the IGM} \label{noescsec}

\subsubsection{Properties of a nebula} \label{nebpropsec}

Numerical simulations suggest that when a \pop\ star forms, the nebula
consists of a higher density phase, with $n_{\mathrm
H}\simeq10^4~\cmt$, and a lower density phase (e.g., Bromm et
al. 1999).  We make the simplifying assumption that half of the
nebula's mass is contained in a homogeneous phase with density
$n_{\mathrm H}=10^4~\cmt$ that completely covers the star(s), and
ignore the lower density gas.  We take the mass fractions of hydrogen
and helium as $X=0.75$ and $Y=0.25$, respectively.

Ionizing radiation from the star(s) creates an \hii\ region in the
dense nebula.  Because of the hardness of our input spectrum, in the
inner part of the \hii\ region helium is doubly ionized (the \heiii\
region).  In the outer part it is singly ionized (the \heii\ region;
the spectrum is hard enough that there is no \hii/\hei\ region).  The
majority of photons above the \heii\ ionization threshold ionize
\heii\ rather than \hi.  We iteratively solve for the sizes and
temperatures of these regions, using the thermodynamic equations from
Cen \shortcite{cen92}.  The \heiii\ region, comprising $0.4$ of the
volume of the \hii\ region, has a temperature of
$3.6\times10^4~\kelvin$; the \heii\ region is cooler, at
$2.7\times10^4~\kelvin$.  In both regions the primary cooling
mechanism is \hi\ recombination, but free-free emission and cooling
via collisional excitation of \hi\ are also important.  The total
nebular emission is not very sensitive to the relative sizes of the
\heii\ and \heiii\ regions.

Given these properties of an ionized region, we can determine
$f_{\mathrm esc}$.\footnote{The dynamical evolution of the gas on the
timescale of the lifetime of a \pop\ star may be important, as the
pressure of the nebular gas may exceed the gravitational pressure, but
here we treat the nebula as static.} The volume and mass of the
\heiii\ region are
\begin{equation}
V_{\mathrm HeIII} = 
\frac{Q_{\mathrm HeII}} 
{n_{\mathrm e}n_{\mathrm HeIII}
\alpha_{\mathrm B}(\heii,T=3.6\times10^4~\kelvin)},
\end{equation}
\begin{equation}
M_{\mathrm HeIII} = \mu\,m_{\mathrm H}
\left(n_{\mathrm H}+n_{\mathrm He}\right)
V_{\mathrm HeIII}
\simeq 0.7 \left(\frac{M_*}{\msun}\right),
\end{equation}
where $Q_{\mathrm HeII}$ is the stellar emission rate of photons
energetic enough to ionize \heii, $\alpha_{\mathrm B}(\heii,T)$ is the
Case B recombination coefficient for \heii, here $M_*$ is the mass of
the ionizing \pop\ star(s) in the nebula and $\mu=1.2$ is the mean
molecular weight.  Because in the \heiii\ region recombinations to
\heii\ provide enough photons to keep the hydrogen ionized, the volume
and mass of the \heii\ region are
\begin{equation}
V_{\mathrm HeII} =
\frac{Q_{\mathrm HI}-Q_{\mathrm HeII}}
{n_{\mathrm e}n_{\mathrm HII}
\alpha_{\mathrm B}(\hi,T=2.7\times10^4~\kelvin)},
\end{equation}
\begin{equation}
M_{\mathrm HeII} = 
\mu\,m_{\mathrm H}\left(n_{\mathrm H}+n_{\mathrm He}\right)
V_{\mathrm HeII}
\simeq 1.0 \left(\frac{M_*}{\msun}\right),
\end{equation}
where $Q_{\mathrm HI}$ is the rate of \hi-ionizing photons, and
$\alpha_{\mathrm B}(\hi,T)$ is the hydrogen recombination coefficient.

The total mass of ionized gas is then $1.7$ times the mass of the
ionizing star(s).  In the single-burst star-formation model, the
nebula mass in $n_{\mathrm H}=10^4~\cmt$ gas, $M_{\mathrm neb}$, is
\begin{equation}
M_{\mathrm neb} = \frac{1}{2}\,\frac{1-\eta}{\eta}\,M_*.
\end{equation}
For $\eta=0.4$, $M_{\mathrm neb}=0.75\,M_*$, which implies $f_{\mathrm
esc}=0$ would be difficult to achieve for our model nebula.  In the
ongoing model, though, $M_{\mathrm neb}>(1-\eta)M_*/(2\eta)$ in
general, and especially at lower redshift.  This is due to the short
lifetime of \pop\ stars: star formation in a halo subsequent to the
first episode usually takes place after previous generations of stars
in the halo have stopped radiating.  Thus $f_{\mathrm esc}=0$ may be
possible in these haloes.

\subsubsection{Resulting spectrum}

\begin{figure}
\psfig{file=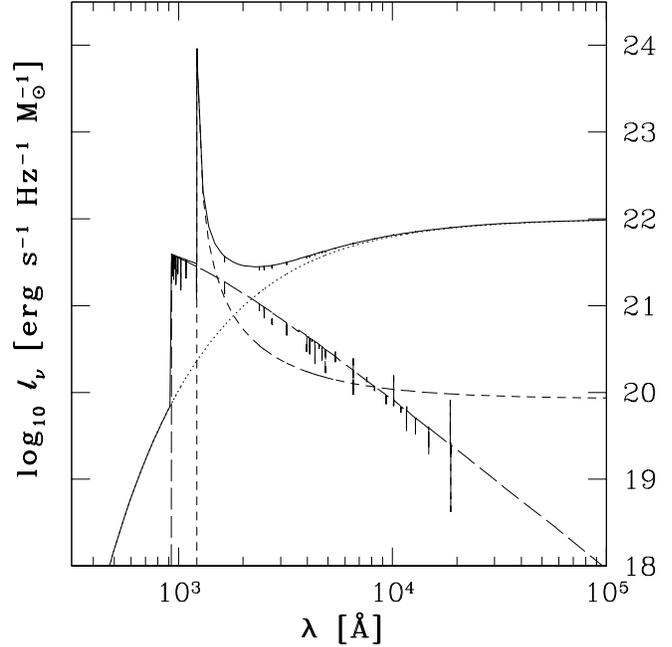,width=3.4in}
\caption{Spectrum of a $z=15$ \pop\ star plus nebular emission, for
the $f_{\mathrm esc}=0$ case of section \ref{noescsec}.  The
long-dashed line is the spectrum of the star, cut off for absorption
shortward of the Lyman limit.  The dotted line is the spectrum of
free-free emission from the nebula.  The short-dashed line is emission
from $\lya$ recombination in the nebula at $z=15$, corrected for
scattering in the IGM.  The solid line is the sum of the spectra.  All
spectra are in the rest frame of the star.}
\label{procspec15fig}
\end{figure}

\paragraph{Recombination emission.} \label{recemsec}
Approximately $1/2$ of the energy radiated by a \pop\ star in a nebula
is ultimately reradiated by recombinations.  The mean energy of a free
electron just before it recombines with a proton may be estimated from
\begin{equation}
\left<E\right> = 
\frac{\beta_{\mathrm B}(\hi,T)\, k T}{\alpha_{\mathrm B}(\hi,T)},
\end{equation}
where $\beta_{\mathrm B}(\hi,T)$ is the recombination emission
coefficient.  Using $T=3\times10^4~\kelvin$, we find
$\left<E\right>=1.4~\ev$ \cite{sea59,cen92}.  The average energy
radiated per recombination, then, is $15.0~\ev$.  Under the
on-the-spot approximation, all Lyman series photons degrade into
$n=2\rightarrow1$ transitions, where $n$ is the energy level of an
excited hydrogen atom.  Since $10.2~\ev$ is released per
$n=2\rightarrow1$ transition, about $2/3$ of the recombination energy
is released in $\lya$ or two-photon emission, and $1/3$ of the
recombination energy (and about $1/6$ of the total emitted energy) is
emitted in other recombination lines and free-bound continuous
emission.  We ignore those processes here; see Schaerer
\shortcite{sch02} for a more complete computation of nebular emission
lines and free-bound spectra.

The relative importance of $\lya$ emission compared to two-photon
emission is determined by the effective recombination coefficients to
the $2p$ and $2s$ states, respectively, and the collisional excitation
rate from the $2s$ state to the $2p$ state.  At
$T=3\times10^4~\kelvin$, $0.75$ $\lya$ photons are emitted for every
\hi\ recombination \cite{ost89,sto95}; we ignore the contribution of
two-photon emission (about 12 per cent of the total emission) to the
spectrum.

Since $\lya$ photons resonantly scatter in neutral hydrogen, they will
not travel far in the IGM until their frequencies are shifted away
from the resonant frequency, $\nu_\lya$.  Photons initially scattered
blueward of the line resonance will eventually cosmologically redshift
back into the resonance.  The result is that the IGM ultimately
scatters all $\lya$ photons to the red side of the line resonance
(broadening from motions inside a halo doesn't contribute to the line
profile).  For a homogeneous, expanding IGM, the resulting scattered
line profile, $\phi(\nu,z)$, was simulated by Loeb \& Rybicki
\shortcite{loe99}, and we fit their result with
\begin{equation}
\phi(\nu,z)=\left\{ 
\begin{array}{ll}
\nu_{*}(z)\,\nu^{-2} 
\exp\left[\frac{-\nu_{*}(z)}{\nu}\right] & \mathrm{if~}\nu>0 \\
0 & \mathrm{if~}\nu\leq0\\
\end{array}\right., 
\end{equation}
\begin{equation}
\nu_{*}(z) = 
1.5\times10^{11}~\hz~\left(\frac{\Omega_{\mathrm B} h^2}{0.019}\right)
\left(\frac{h}{0.7}\right)^{-1}
 \frac{(1+z)^3}{E(z)}.
\end{equation}
This profile results in a strong, asymmetric $\lya$ emission line near
$1220-1225~\ang$ with a scattering tail extending to long wavelengths.

In the \heiii\ region, \heii\ recombinations sometimes produce more
than one photon capable of ionizing \hi.  Recombinations directly to
the $n=2$ state produce \heii\ Balmer continuum photons, which are
capable of ionizing \hi.  In addition, two-photon decay from the $2s$
state produces 1.42 \hi-ionizing photons per decay \cite{ost89}.
Since \heii\ $\lya$ is also capable of ionizing hydrogen, the mean
number of hydrogen ionizations per \heii\ recombination is
\begin{equation}
\frac{
\alpha_2(\heii,T)+1.42\alpha_{2s}^{\mathrm eff}(\heii,T)+
\alpha_{2p}^{\mathrm eff}(\heii,T)
}
{\alpha_{\mathrm B}(\heii,T)}=1.7,
\end{equation}
where $\alpha_2(\heii,T)$ is the recombination coefficient for
recombinations directly to the $n=2$ state and $\alpha_{2l}^{\mathrm
eff}(\heii,T)$ is the effective total recombination coefficient to the
$2l$ state \cite{sto95}.

\paragraph{Free-free emission.}
Free-free radiation accounts for about $1/4$ of the cooling in the
nebula.  This energy is radiated in a continuous spectrum,
\begin{displaymath}
j_\nu^{\mathrm ff} = 
7.2\times10^{-39}\,\sum_{Z} n_{\mathrm e} n_{Z}\,
Z^2\left(\frac{T}{\kelvin}\right)^{-1/2}
\end{displaymath}
\begin{equation}
\hspace{2 cm}\times\exp\left(\frac{-h\,\nu}{k\,T}\right)
~\mathrm{erg~s^{-1}cm^3\,Hz^{-1}ster^{-1}},
\end{equation}
where $j_\nu^{\mathrm ff}$ is the specific emission coefficient, $\nu$
is the frequency of emitted radiation and $n_Z$ is the number density
of ions of net charge $Z$ \cite{fer80} A Gaunt factor of 1.3 has been
assumed; this results in an error of less than 8 per cent over optical
and UV frequencies \cite{kar61}.

The luminosity per solar mass of \pop\ stars from free-free radiation,
$l_\nu^{\mathrm ff}$, is
\begin{equation} \label{ffspeceq}
l_\nu^{\mathrm ff} = 
\frac{4\pi}{M_*} 
\left( j_\nu^{\mathrm ff,HeIII}~ V_{\mathrm
HeIII} + j_\nu^{\mathrm ff,HeII}~ V_{\mathrm HeII} \right).
\end{equation}
Here \heiii\ and \heii\ label the emission coefficients and volumes
computed for the \heiii\ and \heii\ {\it regions}; in each case the
free-free emission is dominated by \hii.

\paragraph{Emission from collisional excitation of \hi.}
Collisions between a free electron and trace \hi\ atom sometimes
excite the \hi\ atom, which then radiates away the excitation energy.
The collisional excitation rate coefficient for transitions from the
$n=1$ state to state $u$, $q_{1u}$, is
\begin{equation} \label{collrateeq}
q_{1u} = \frac{8.629\times10^{-6}}{T^{1/2}} 
\frac{\Omega(1,u)}{\omega_1}
\exp\left(\frac{-\chi(1,u)}{k\,T}\right)~\se^{-1}\,\cm^3,
\end{equation}
where $T$ is in $\kelvin$, $\Omega(1,u)$ is the
(temperature-dependent) effective collision strength for transitions
from the $n=1$ to state $u$, $\omega_1$ is the statistical weight of
the $n=1$ state and $\chi(1,u)$ is the energy difference between the
$n=1$ and state $u$ \cite{ost89}.

We compute the collisional excitation to the $n=2$ and $n=3$ states
using cross-sections from Callaway \shortcite{cal85} and Callaway,
Unnikrishnan \& Oza \shortcite{cal87}.  These excitations result in
additional $\lya$ emission: excitations to the $2p$, $3s$ and $3d$
states radiatively decay to $n=1$ via $\lya$ photons, and atoms in the
$2s$ state (resulting from collisions to either the $2s$ or $3p$
state) may be additionally collisionally excited to the $2p$ state.
The rate of collisional $\lya$ emission per unit stellar mass,
$q_\lya^{\mathrm coll}$, is
\begin{equation} \label{cereq}
q_\lya^{\mathrm coll} = 
n_{\mathrm e} n_{\mathrm HI}~q_{1,2p}^{\mathrm eff}
\left(\frac{V}{M_*}\right), 
\end{equation}
where 
\begin{equation}
q_{1,2p}^{\mathrm eff} = 
\sum_u L(u) \, q_{1u}
\end{equation}
and
\begin{equation}
L(u)=\left\{ 
\begin{array}{ll}
1 & {\mathrm if}~u\in(2p,3s,3d) \\
0.33 & {\mathrm if}~u\in(2s,3p)\\
\end{array}\right.. 
\end{equation}
The factor of 0.33 accounts for $2s$ to $2p$ collisional excitation
\cite{ost89}.  Summing over the \heii\ and \heiii\ regions,
$q_\lya^{\mathrm
coll}=3.4\times10^{47}~\se^{-1}\,\mathrm{M^{-1}_{\odot}}$.

\paragraph{Total spectrum.}
For the $f_{\mathrm esc}=0$ case, the total specific luminosity per
unit stellar mass emitted from a \pop\ star and nebula and scattered
in the IGM, $l_{\nu}^0(z)$, has three components that we treat: the
truncated stellar spectrum, $l_{\nu}^{*{\mathrm a}}$, the free-free
spectrum, $l_{\nu}^{{\mathrm ff}}$, and the scattered $\lya$ spectrum,
$l_{\nu}^{\lya}(z)$:
\begin{equation}
l_{\nu}^0(z) = 
l_{\nu}^{*{\mathrm a}} + l_{\nu}^{{\mathrm ff}} + l_{\nu}^{\lya}(z).
\end{equation}
The truncated stellar spectrum is
\begin{equation} \label{trspeceq}
l_{\nu}^{*{\mathrm a}} = \left\{ 
\begin{array}{ll}
l_{\nu}^{*} & \mathrm{if~}h\nu<13.6~\ev \\
0 & \mathrm{if~}h\nu\geq13.6~\ev \\
\end{array}\right..
\end{equation}
The $l_{\nu}^{*{\mathrm a}}$ spectrum is slightly modified between
$912$ and $1216~\ang$ by scattering in the IGM, when the photons
in that range are cosmologically redshifted into the $\lya$ resonance
(e.g., Peebles 1993).  This effect is small when
$1.75\,\nu_{*}(z)\ll8.22\times10^{14}~\hz$, i.e., the width of the
scattered $\lya$ line \cite{loe99} is small compared to the frequency
difference between $912~\ang$ and $1216~\ang$.  Since
$1.75\,\nu_{*}(z=30)=8.27\times10^{13}~\hz$, we ignore this correction
to the spectrum.

The free-free spectrum is given by eq. (\ref{ffspeceq}).  The $\lya$
spectrum is
\begin{equation} \label{lyaspeceq}
l_{\nu}^{\lya}(z) = q_\lya ~
h\nu_{\lya}~\phi(\nu_{\lya}-{\nu},z),
\end{equation}
where $q_\lya$ is the rate of $\lya$ photons produced per solar mass
of the ionizing star.  That rate is
\begin{equation}
q_\lya = 
0.75\left(q_{\mathrm HI}-q_{\mathrm HeII}
+1.7\,q_{\mathrm HeII}\right)+q_\lya^{\mathrm coll},
\end{equation}
where $q_{\mathrm HI}\equiv Q_{\mathrm HI}/M_*$ and $q_{\mathrm
HeII}\equiv Q_{\mathrm HeII}/M_*$.  The factor $0.75$ represents the
fraction of hydrogen recombinations that result in $\lya$ photons, and
the factor of $1.7$ accounts for the number of hydrogen ionization per
\heii\ recombination, computed above.  

\mbox{Figure \ref{procspec15fig}} show $l_{\nu}^{*{\mathrm a}}$,
$l_{\nu}^{{\mathrm ff}}$, $l_{\nu}^{\lya}(z)$ and $l_{\nu}^0(z)$ for
$z=15$.  In this model no \pop\ ionizing photons escape to the IGM,
thus \pop\ stars don't contribute to reionization of the Universe
(cf. section \ref{allescsec}).

\subsection{Complete escape of ionizing radiation into the IGM} 
\label{allescsec}

\subsubsection{Properties of the IGM}

\begin{figure}
\psfig{file=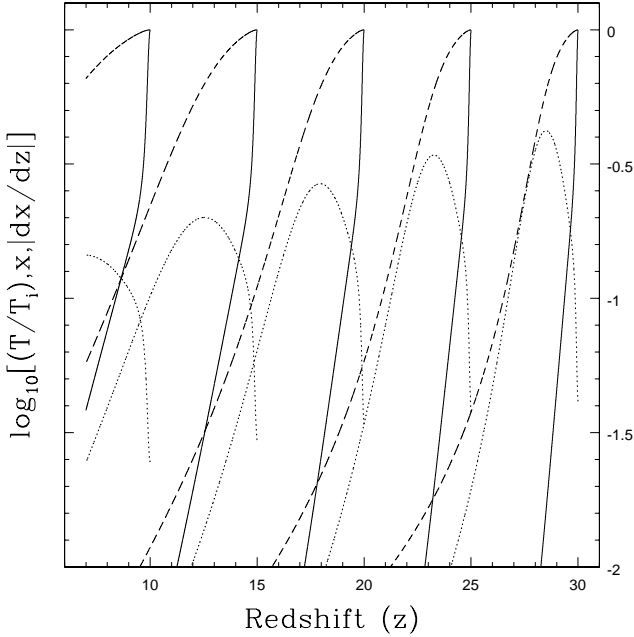,width=3.4in}
\caption{Properties of the ionized IGM near a \pop\ star.  There are
sets of three curves for five values of $z_{\mathrm
i}=(10,15,20,25,30)$.  In each set of curves, the solid line is the
temperature as a fraction of the initial temperature, the dashed line
is the ionization fraction and the dotted line is $|{\dd x}/{\dd z}|$,
followed from $z_{\mathrm i}$ to $z=7$.}
\label{igmplotfig}
\end{figure}

The second case we consider is that the nebula plays no role in
reprocessing ionizing radiation from a \pop\ star, $f_{\mathrm
esc}=1$.  This may be because, in contrast to the assumptions we made
in \S \ref{nebpropsec}, the density of the nebula is low, or because
the nebula is clumped into high-density regions with a small covering
fraction, or because the nebula was blown away by the star(s).
Because the timescales of important IGM processes extend beyond the
lifetime of a \pop\ star, in this section we will name the redshift of
formation of a \pop\ star $z_{\mathrm i}$, and then describe the
evolution as a function of $z$.

We assume the IGM is uniform with baryon density $n_{\mathrm IGM}(z)=
1.7\times10^{-7}(1+z)^3~\cmt$ (which ignores the small fraction of
baryons in collapsed haloes), $X=0.75$ and $Y=0.25$.  Ionizing photons
from a \pop\ star stream into the IGM and form an ionized region.
Because the density is low, for $z_{\mathrm i}\la30$ recombinations
are of little importance on the timescale of the star's lifetime,
$\tau \simeq 2\times10^6~\yr$.  For the purposes of calculating the
properties of the ionized region of the IGM, we assume that all of the
ionizations occur immediately; at worst this contributes less than a 3
per cent error to the computed spectrum.

\heiii\ recombines more quickly than \hii, and \hei\ recombinations
occur on a comparable timescale to those of \hi.  As discussed in
\S\ref{recemsec}, \heii\ recombinations produce $1.7$ \hi-ionizing
photons on average.  In the low-density limit, a \hei\ recombination
ionizes $1$ hydrogen atom \cite{ost89}.  We treat \hei\ and \heii\
ionizations as an extra $1$ and $1.7$ \hi\ ionizations, respectively,
and track only hydrogen recombinations.

For $z_{\mathrm i}\la30$, a \pop\ star ionizes a volume of the IGM,
$V(z)$, initially given by
\begin{equation} \label{littleveq}
V(z_{\mathrm i}) = 
\frac{\left(Q_{\mathrm HI}+0.7\,Q_{\mathrm HeII}\right) \tau}
{n_{\mathrm IGM}(z)} = 
\frac{2.4\times10^{4}}{(1+z_{\mathrm i})^{3}}
\left(\frac{M_*}{\msun}\right)
\mathrm{kpc^3}.
\end{equation}
The initial temperature in this volume is determined from
photoionization heating to be $T_{\mathrm i}\simeq
6\times10^4~\kelvin$.  Given $z_{\mathrm i}$, we solve the temperature
evolution of the ionized IGM region as a function of $z$ until $z=7$,
when we assume that the Universe reionizes.  If \pop\ stars reionize
the Universe before $z=7$, their ionized nebulae will have overlapped,
and the analysis of this section will no longer be appropriate.  The
post-reionization contribution of the IGM to the CIRB depends on the
temperature and ionization state of the IGM, but is expected to be
small under typical assumptions.  But the most important effect that
reionization would have on the \pop\ contribution the CIRB may be the
effect reionization has on the SFR.  As noted in \S\ref{sfrmodsec},
reionization may stop or severely curtail the formation of \pop\
stars.

To compute the temperature of the ionized IGM around a \pop\ star as a
function of redshift, we set the initial temperature to $T_{\mathrm
i}=6\times10^4~\kelvin$ and follow the evolution of temperature
considering cooling from Compton scattering of CMBR photons off of
electrons, adiabatic cooling of the IGM from the expansion of the
Universe, cooling from collisional excitation of \hi, free-free
cooling and recombination cooling \cite{cen92}.  Since all of those
processes except adiabatic cooling depend on the ionization fraction,
$x$, it is solved for simultaneously, using the approximation
$x\simeq{n_{\mathrm e}}/{n_{\mathrm IGM}}$:
\begin{equation}
-\frac{\dd x}{\dd t} =
\alpha_{\mathrm B}(\hi,T) {n_{\mathrm IGM}} x^2
\end{equation}
(e.g., Peebles 1993).  Figure \ref{igmplotfig} plots $T(z)$, $x(z)$
and $|{\dd x(z)}/{\dd z}|$ for five values of $z_{\mathrm i}$.

Initially nearly all the hydrogen is ionized, and Compton cooling
briefly dominates the thermodynamics.  Once a small fraction,
$\sim10^{-4}$, of the hydrogen atoms have recombined, cooling via
collisional excitation becomes the dominant process.  As the
temperature decreases, the recombination rate increases, which in turn
increases the neutral fraction and thus the collisional cooling rate.
When the temperature reaches about $1.5\times10^4~\kelvin$, Compton
cooling dominates again.  For lower values of $z_{\mathrm i}$,
adiabatic cooling eventually dominates; for higher values of
$z_{\mathrm i}$, recombination cooling becomes important.  Free-free
emission is important enough to be included.

\subsubsection{Resulting spectrum}
\label{allescspecsec}

\begin{figure}
\psfig{file=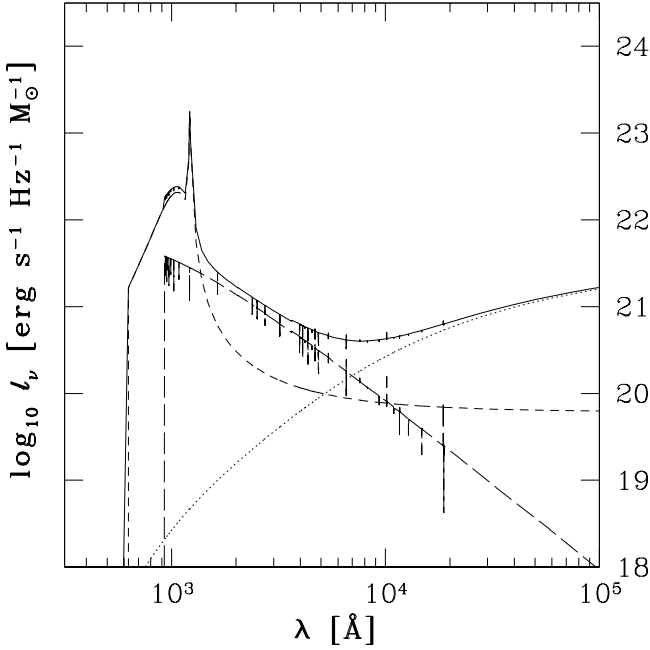,width=3.4in}
\caption{Spectrum of a $z=15$ \pop\ star plus emission from the IGM,
for the $f_{\mathrm esc}=1$ case of section \ref{allescsec}.  The
long-dashed line is the spectrum of the star, cut off for absorption
shortward of the Lyman limit.  The dotted line is the spectrum of
free-free emission from the IGM.  The short-dashed line is emission
from $\lya$ recombination in the IGM, corrected for scattering.  The
free-free and $\lya$ spectra are integrated from $z=15$ to $z=7$, and
divided by the lifetime of a \pop\ star to give a useful
normalization.  The solid line is the sum of the spectra.  The sharp
peak at $1216~\ang$ is collisionally excited $\lya$ emission; the
broad peak at $1000~\ang$ is $\lya$ from recombinations.  All spectra
are in the rest frame of the star (see Section \ref{allescspecsec} for
explanation).}
\label{igmspec15fig}
\end{figure}

\paragraph{Recombination emission.}
As in \S\ref{recemsec}, we consider only the $\lya$ component of the
recombination spectrum.  The rate of production of $\lya$ photons from
recombinations, $q_\lya^{\mathrm rec}$, is
\begin{equation}
q_\lya^{\mathrm rec} = 0.63\left(q_{\mathrm HI}-q_{\mathrm HeII}
+1.7\,q_{\mathrm HeII}\right)\tau
\left|\frac{\dd x}{\dd t}\right|,
\end{equation}
where $0.63$ is a representative value for the fraction of hydrogen
recombinations resulting in $\lya$ emission near the peak of the
recombination rate.

\paragraph{Free-free emission.}
The free-free spectrum is
\begin{displaymath}
j_\nu^{\mathrm ff} = 
7.2\times10^{-39}\,x^2\,n_{\mathrm IGM}^2\,
\left(\frac{T}{\kelvin}\right)^{-1/2}
\end{displaymath}
\begin{equation}
\hspace{2 cm}\times\exp\left(\frac{-h\,\nu}{k\,T}\right)
~\mathrm{erg~s^{-1}cm^3\,Hz^{-1}ster^{-1}}
\end{equation}
\cite{fer80}.  The specific luminosity per unit stellar mass is
\begin{displaymath}
l_\nu^{\mathrm ff} (z,z_{\mathrm i}) = 
4\pi\,j_\nu^{\mathrm ff}\,\frac{V(z)}{M_*} =
1.6\times10^{18}\,x^2\,(1+z)^3
\left(\frac{T}{\kelvin}\right)^{-1/2}
\end{displaymath}
\begin{equation} \label{igmffeq}
\hspace{2 cm}\times\exp\left(\frac{-h\,\nu}{k\,T}\right)
~\mathrm{erg~s^{-1}\,Hz^{-1}\mathrm{M}_\odot^{-1}},
\end{equation}
where the second equality follows from equation (\ref{littleveq}).
The $z_{\mathrm i}$ dependence of $l_\nu^{\mathrm ff}$ results from
the implicit dependence of $T$ and $x$ on $z_{\mathrm i}$ and $z$.
 
\paragraph{Emission from collisional excitation of \hi.}

The rate of $\lya$ photons produced by collisional excitations is
given by eqs. (\ref{cereq}) and (\ref{littleveq}):
\begin{equation}
q_\lya^{\mathrm coll} = 
2.0\times10^{55} ~x \, (1-x)~
(1+z)^{3} ~q_{1,2p}^{\mathrm eff}
~\mathrm{cm^{-3}~M_\odot^{-1}}.
\end{equation}
In the low density
limit, though, collisions from the $2s$ to $2p$ state are unimportant,
so
\begin{equation}
L(u)=\left\{ 
\begin{array}{ll}
1 & {\mathrm if}~u\in(2p,3s,3d) \\
0 & {\mathrm if}~u\in(2s,3p)\\
\end{array}\right.. 
\end{equation}
Additionally, because the temperature in the IGM varies, we make the
following fits to the data of Callaway et al. (1987):
\begin{equation}
\Omega(1,2p)\simeq -2.41\times10^{-3}\left(\frac{T}{10^4}\right)^2
+0.148\left(\frac{T}{10^4}\right)+0.170
\end{equation}
and
\begin{displaymath}
\Omega(1,3s)+\Omega(1,3d) 
\end{displaymath}
\begin{equation}
\hspace{1. cm}\simeq 
-2.29\times10^{-3}\left(\frac{T}{10^4}\right)^2
+0.0299\left(\frac{T}{10^4}\right)+0.116.
\end{equation}
The fits are good to better than 5 per cent for temperatures from
$1.6\times10^4$ to $6\times10^4~\kelvin$, where almost all of the
collisional excitation occurs.

\paragraph{Total spectrum.}
For the $f_{\mathrm esc}=1$ case, the total specific luminosity per
unit stellar mass emitted from a \pop\ star and IGM, including
scattering, is $l_{\nu}^0(z)$.  We treat the same three components as
in the $f_{\mathrm esc}=0$ case: the truncated stellar spectrum,
$l_{\nu}^{*{\mathrm a}}$, the free-free spectrum, $l_{\nu}^{{\mathrm
ff}}(\nu,z,z_{\mathrm i})$, and the scattered $\lya$ spectrum,
$l_{\nu}^{\lya}(\nu,z,z_{\mathrm i})$.  The truncated stellar spectrum
is given by eq. (\ref{trspeceq}), and the free-free spectrum is given
by eq. (\ref{igmffeq}).  Similar to eq. (\ref{lyaspeceq}), the $\lya$
spectrum is
\begin{equation} 
l_{\nu}^{\lya}(z,z_{\mathrm i}) = q_\lya
~ h\nu_{\lya}~\phi(\nu_{\lya}-{\nu},z),
\end{equation}
where
\begin{equation}
q_\lya(z,z_{\mathrm i}) = q_\lya^{\mathrm rec}+q_\lya^{\mathrm coll}.
\end{equation}

The total spectrum emitted by a \pop\ star and the surrounding IGM
from the formation of the star at $z_{\mathrm i}$ to the redshift at
which the Universe is reionized, $z=z_{\mathrm reion}$, expressed in
the rest frame of the star when it forms and normalized to the
lifetime of the star, is
\begin{displaymath}
l_{\nu}^1(\nu,z_{\mathrm i}) = 
l_{\nu}^{*{\mathrm a}}(\nu)
\end{displaymath}
\begin{equation}
\hspace{0.5 cm}+ \frac{1}{\tau} \int^{z_{\mathrm i}}_{z_{\mathrm reion}} 
\dd z ~
\left|\frac{\dd t}{\dd z}\right|
\left[l_{\nu}^{{\mathrm ff}}
(\nu',z,z_{\mathrm i})
+ l_{\nu}^{\lya}
(\nu',z,z_{\mathrm i})
\right],
\end{equation}
with
\begin{equation}
\nu' = \nu  \left(\frac{1+z}{1+z_{\mathrm i}}\right).
\end{equation}
Because this spectrum is in the frame of the star when it forms, IGM
emission at later times (and thus lower values of $z$) appears in the
spectrum at {\it higher} frequencies than it is emitted at; this is
because we {\it blueshift} all of the IGM radiation back to redshift
$z_{\mathrm i}$.  The spectra are normalized by integrating the
emission from $z_{\mathrm i}$ to $z=7$ and dividing by the lifetime of
the star, $\tau$.  The resulting spectrum isn't meaningful for the
observation of a single \pop\ stellar source plus ionized IGM region,
because the photons are emitted at different times, but it is a useful
spectrum for computing the CIRB.  The spectrum of a single source at
redshift $z<z_{\mathrm i}$ (such that the star no longer contributes
directly to the spectrum) is
\begin{equation}
l_{\nu}^1(\nu,z,z_{\mathrm i}) = 
l_{\nu}^{{\mathrm ff}}
(\nu,z,z_{\mathrm i})
+ l_{\nu}^{\lya}
(\nu,z,z_{\mathrm i});
\end{equation}
this spectrum is in the frame of the emission at redshift $z$.

\mbox{Figure \ref{igmspec15fig}} shows $l_{\nu}^{*{\mathrm a}}$,
$l_{\nu}^{{\mathrm ff}}(z_{\mathrm i})$, $l_{\nu}^{\lya}(z_{\mathrm
i})$ and $l_{\nu}^1(z_{\mathrm i})$ for $z_{\mathrm i}=15$.  The sharp
emission peak at $1216~\ang$ results from collisionally-excited $\lya$
emission shortly after the star forms.  The broad peak centered near
$1000~\ang$ is $\lya$ emission from recombinations.  At $608~\ang$ the
$\lya$ spectrum is cut off as a consequence of integrating the IGM
spectrum to $z=7$; the cut-off location is $1216~\ang\,(1+z_{\mathrm
reion})/(1+z_{\mathrm i})$.

\section{Cosmic infrared background} \label{cirbsec}

\begin{figure}
\psfig{file=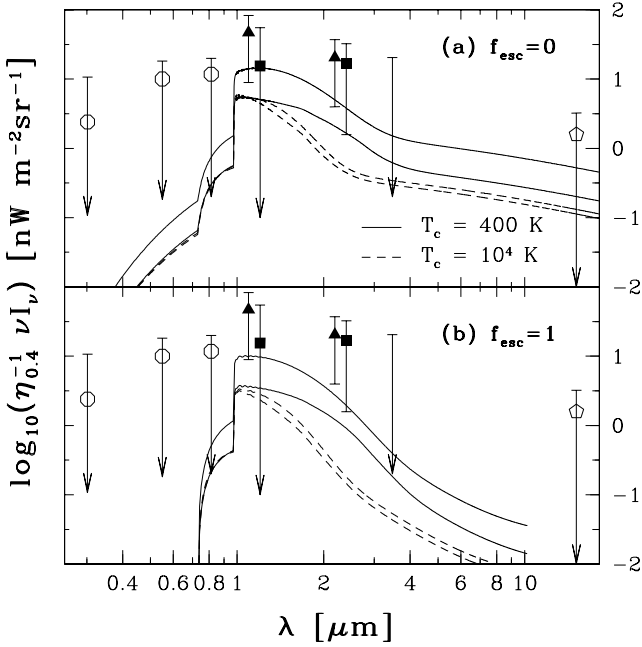,width=3.4in}
\caption{The cosmic infrared background from \pop\ stars.  The
ordinate is observed frequency multiplied by observed specific
intensity.  Shown are curves for star formation in haloes with virial
temperatures above critical temperatures $\tcrit=400~\kelvin$
(molecular-hydrogen cooling, solid lines) and $10^4~\kelvin$
(atomic-hydrogen cooling, dashed lines).  (a) $f_{\mathrm esc}=0$.
(b) $f_{\mathrm esc}=1$.  For both $f_{\mathrm esc}$ cases, there are
two sets of two curves: the upper set is for the ongoing
star-formation model; the lower set is for the single-burst model.
The properties of the curves are explained in \S4.1; this figure shows
$z_{\rm end}=7$.  The points show the excess CIRB, with 2$\sigma$
errors, and are described in \S4.2.}
\label{cirbfig}
\end{figure}

\subsection{Model results}
We now convolve the processed spectra of Section \ref{specsec} with
the star-formation histories of Section \ref{sfrsec} to calculate the
\pop\ contribution to the CIRB.  The background is evaluated with
\begin{equation} \label{cirbcalceq}
\nu_{\mathrm obs}I_{\nu} = 
{c}\int^{\infty}_{0}~{\dd}z~
\left|\frac{{\dd}t}{{\dd}z}\right|\frac{\nu(z) j^{\mathrm c}_{\nu}(z)}{1+z}
\end{equation}
(e.g., Peebles 1993).  Here $\nu_{\mathrm obs}$ is the observed
frequency, $I_{\nu}$ is the observed specific intensity,
$\nu(z)=(1+z)\nu_{\mathrm obs}$ and $j^{\mathrm c}_{\nu}(z)$ is the
comoving specific emission coefficient.  Both the star-formation
history and the assumed recombination history determine $j^{\mathrm
c}_{\nu}(z)$:
\begin{equation} \label{comovej}
j^{\mathrm c}_{\nu}(z) = \frac{1}{4\pi}l_{\nu} \tau \psi(z),
\end{equation}
where $l_{\nu}$ is either $l_{\nu}^{0}$ or $l_{\nu}^{1}$,
$\tau=2\times 10^6~\yr$ is the fiducial main-sequence lifetime of
a \pop\ star \cite{bro01c} and $\psi(z)$ is either $\psi_{\mathrm
on}(z)$ or $\psi_{\mathrm burst}(z)$, the SFR per comoving volume.

Results for $f_{\mathrm esc}=0$ and $1$ are shown in
\mbox{Fig. \ref{cirbfig}}.  There are a total of eight curves in the
figure, generated by varying each of three parameters over two values:
$f_{\mathrm esc}=0$ or $1$; $\tcrit=400$ or $10^4~\kelvin$;
star-formation mode of ongoing or single-burst.  Models with
$f_{\mathrm esc}=0$ always produce more CIRB at all wavelengths than
$f_{\mathrm esc}=1$ models.  Models with $\tcrit=400~\kelvin$ produce
more CIRB than $\tcrit=10^4~\kelvin$ models at all wavelengths except
near $1~\mu$m for single-burst star-formation models.  The ongoing
star-formation models produce more CIRB at all wavelengths than
single-burst star-formation models, except near $1~\mu$m for the
$f_{\mathrm esc}=0$ case.

The sharp edge at $1~\mu$m in all of the curves is a result of our
sharp truncation of \pop\ star formation at $z_{\rm end}=7$.  The edge
occurs at the redshifted wavelength of $\lya$ from stars at $z_{\rm
end}$, i.e., $(1+z_{\rm end})1216$~\AA.  The reason that the CIRB
curves for $\tcrit=400$ and $10^4~\kelvin$ have similar values at
$1~\mu$m for single-burst star-formation models is that the
corresponding SFRs at $z=7$ are similar (see Fig. \ref{sfretafig}).

The ``bump'' in the CIRB curves from $0.7~\mu$m to $1~\mu$m results
from stellar emission between rest-frame 912 and 1216~\AA\ (see
Figs. \ref{procspec15fig} and \ref{igmspec15fig}) by sources at $z=7$.
From $\lambda=1~\mu$m to roughly $2~\mu$m the \pop\ CIRB is dominated
by $\lya$ emission from sources at $1+z=\lambda/1216~\ang$.  The
inflection points of the curves, particularly clear in the $f_{\mathrm
esc}=0$ panel, occur at the wavelength where the \pop\ CIRB
transitions from being dominated by $\lya$ emission to being dominated
by continuum emission from sources at $z=z_{\rm end}$.  The continuum
radiation for the $f_{\mathrm esc}=0$ spectrum is dominated by
free-free emission (see Fig. \ref{procspec15fig}), whereas stellar
continuum is important as well for the $f_{\mathrm esc}=1$ continuum
spectrum (see Fig. \ref{igmspec15fig}).  Consequently, the \pop\
mid-IR background is determined predominantly by the choice of $z_{\rm
end}$ rather than the SFR at extremely high redshifts.  The \pop\
contribution to the optical background (for $z_{\rm end}\ge7$) is due
to the stellar ``bump'' described above and the high-energy tail of
free-free emission, which is only significant for the $f_{\mathrm
esc}=0$ spectrum; this is a consequence of the temperature of the gas,
which is effectively much higher in the $f_{\mathrm esc}=0$ case than
for the $f_{\mathrm esc}=1$ case (see \S3).

\subsection{Observational data}
The points with error bars in \mbox{Fig. \ref{cirbfig}} show the
difference between the total extragalactic background and the
extragalactic background due to resolved sources.  At wavelengths,
e.g., $0.3~\umu$m, where the lower limit is an arrow, the contribution
from resolved sources is sufficient to explain \textit{all} of the
background, at the $2\sigma$ level.  At wavelengths where data points
have lower limits plotted, e.g., $2.2~\umu$m, the light from resolved
sources is not enough to account for the measured background.  It is
this unexplained excess, measured by Cambr\'{e}sy et
al. \shortcite{cam01} at $1.25~\umu$m and by Cambr\'{e}sy et
al. \shortcite{cam01} and Wright \& Johnson \shortcite{wri01} at
$2.2~\umu$m, that we are trying to fit with our models, while
conforming to upper limits at other wavelenths.

\paragraph*{Near-IR data points}
The triangles at 1.25 and $2.2~\umu$m ($J$ and $K$ band, respectively)
correspond to the unexplained CIRB as measured by Cambr\'{e}sy et
al. \shortcite{cam01}: they measured the total extragalactic
background with $COBE$/DIRBE data, and subtracted from that the
contribution due to resolved galaxies measured by near-IR surveys.
The squares at $J$ and $K$ (offset slightly in wavelength for clarity)
are our computation of the unexplained CIRB from the CIRB measurement
of Wright \& Johnson \shortcite{wri01}, using their measurement of the
total extragalactic background from $COBE$/DIRBE data (but using a
different foreground model from Cambr\'{e}sy et al. 2001), and the
same galaxy contribution subtraction and error propagation as
Cambr\'{e}sy et al. \shortcite{cam01}.  The Wright \& Johnson
\shortcite{wri01} point at $J$ band is consistent at the $2\sigma$
level with no unexplained extragalactic background.  All error bars
are $2\sigma$.  The $2\sigma$ upper limit at $3.5~\umu$m ($L$ band)
represents a measurement of the total extragalactic background at that
wavelength \cite{wri01}.  {\it SIRTF} is expected to accurately
measure the contribution of galaxies to the $L$-band CIRB, and thus
allow a determination of the excess $L$-band CIRB (W. T. Reach,
private communication).

\paragraph*{Other data points}
The upper limits at 0.3, 0.6, and $0.8~\umu$m are $2\sigma$ upper
limits on the unexplained optical extragalactic background light
(Bernstein, Freedman, \& Madore 2002).  The open circle data points
show the estimated unexplained optical extragalactic background, which
is consistent with zero ($2\sigma$) at all three bandpasses.
Bernstein et al. (2002) measured the total optical extragalactic
background from $HST$ images (utilising simultaneous ground-based
spectra for absolute zodiacal-light subtraction calibration).  From
published number counts and their own careful photometry of the Hubble
Deep Field observations, they find no evidence for convergence in the
integrated light from galaxies; this suggests that deeper galaxy
photometry will lower the upper limits.

The upper limit at $15~\umu$m is our computation of the $2\sigma$
upper limit on the unexplained mid-IR background light, using the CIRB
upper limits obtained by Renault et al. (2001) from gamma-ray
observations, and the $ISO$/ISOCAM $15~\umu$m galaxy counts reported
by Elbaz et al. (2002).  The open hexagonal data point shows the
estimated excess CIRB, which is consistent with zero ($2\sigma$).

\section{Discussion} \label{discsec}

Our models identify a very narrow range of parameter space in which
\pop\ stars may explain the excess CIRB:
\begin{enumerate}
\item{The \pop\ stars must be very massive, so that most of their
energy is radiated in photons energetic enough to ionize hydrogen.}
\item{Cooling in low-mass haloes is possible due to H$_2$.}
\item{Approximately 40 per cent of the eligible baryons in the star
forming halo must be converted into \pop\ stars.}
\item{The escape fraction of ionizing photons from the nebula
surrounding a \pop\ star must be near zero, so that most of the
ionizing photons are converted into $\lya$ photons.}
\item{\pop\ star formation must begin by $z\simeq25$ and persist until
$z_{\mathrm end}\simeq7$, so that the $\lya$ emission
(Fig. \ref{cirbfig}) extends through the observed $J$ and $K$ bands.}
\item{Negative feedback effects must not inhibit \pop\ star formation.
In particular, we require that \pop\ stars do not radiatively or
mechanically destroy star-forming material, and that they they do not
enrich their surroundings with a sufficient amount of metals to end
\pop\ star formation.}
\end{enumerate}

The amplitudes of the curves in \hbox{Fig. \ref{cirbfig}} scale simply
with $\eta$, and they do not change dramatically with small changes to
the cosmological parameters.  The characteristic breaks in the spectra
are located at $912~\ang(1+z_{\rm end})$ and $1216~\ang(1+z_{\rm
end})$, where $z_{\rm end}$ is the low-redshift limit to \pop\ star
formation; if $z_{\rm end}$ were much higher than 7, the \pop\ CIRB
peak would lie longward of the $J$ band.

If all of the above conditions are met, then {\it all} of the observed
near-IR CIRB deficit is due to Pop III star formation.  Future
observations may demonstrate that other sources contribute
significantly to the near-IR background; e.g., part of the deficit
could be due to the faint wings of galaxies that are unaccounted for
in current surveys (Totani et al. 2001).  The parameter space explored
in this paper easily accomodates lower values of the unexplained CIRB,
through decreasing the efficiency, $\eta$.  The shape of the \pop\
CIRB spectrum in the near-IR is primarily determined by the shape of
the star-formation rate as a function of redshift.

The formation of very massive stars probably requires that the
star-forming gas is not enriched with heavy elements to a mass
fraction $Z_{\mathrm crit}\ga 10^{-3}Z_{\odot}$, where $Z_{\odot}$ is
the solar value (Bromm et al. 2001a). We require a large fraction of
the mass processed through \pop\ stars to end up in massive black
holes, so as not to overproduce metals compared to the observed levels
in the high redshift IGM and to avoid enriching the primordial gas too
quickly to $Z_{\mathrm crit}$. Recently, Schneider et al. (2002) have
pointed out the possible problem that if {\it all} \pop\ stars
collapsed into massive black holes, the IGM would always have zero
metallicity.  It seems natural to assume, however, that a small
fraction of the \pop\ stellar mass, $\Omega_{Z}=Z \Omega_{\mathrm
B}=\epsilon_{\mathrm pi} \Omega_{\mathrm III}$ is ejected into the IGM
through pair-instability supernovae, even if the majority of the mass
is permanently locked up in black holes.  Assuming a ratio
$\Omega_{\mathrm III}/\Omega_{\mathrm B}\sim 0.1$, the required
pair-instability fraction to produce the critical level of metallicity
$Z_{\mathrm crit}$ would be $\epsilon_{\mathrm pi}\sim 10^{-2}$
(see also Oh et al. 2001).

We thus conclude that if \pop\ stars explain a large fraction of the
near-IR background, then almost all \pop\ material must end up in
massive black holes.  
In that case we would predict a
significant contribution, $\sim10$ per cent, by massive black holes to
the total baryonic mass budget in the Universe.  Although such a
prediction may seem somewhat extreme, it will ultimately be tested by
observations (e.g., Agol \& Kamionkowski 2001; Schneider et al. 2002).
We emphasize that at
present there are no observations that rule out such a scenario (e.g.,
Carr 1994).
Moreover, there are good physical reasons to seriously
consider the existence of massive black holes resulting from \pop\
star formation (e.g., Bromm et al. 2002; Madau \& Rees 2001; 
Schneider et al. 2002).

In \S \ref{omegasec} we discussed $\Omega_{\mathrm III}$, the
cumulative matter density processed through \pop\ stars.  In
\hbox{Fig. \ref{omegafig}} the horizontal lines labelled \hi, \hei\
and \heii\, correspond to the $\Omega_{\mathrm III}$ required to
produce 10 ionizing photons per particle of each species (Bromm et
al. 2001b).  For the $f_{\mathrm esc}=1$ model, the intersection of
those lines with the $\Omega_{\mathrm III}$ star curves is expected to
be closely related to the redshift of reionization.  For the
$f_{\mathrm esc}=0$ model, though, \pop\ ionizing photons are all
absorbed in the haloes in which they are emitted, and thus \pop\ stars
make no contribution to the reionization of the Universe.  We
discussed above that if \pop\ stars enriched their immediate
environment with a sufficient amount of metals, \pop\ stars would no
longer form there (producing a SFR similar to our single-burst model).
If metals from \pop\ stars efficiently mixed throughout the Universe,
the era of \pop\ stars might therefore come to an end (see Bromm et
al. 2001a).

A single \pop\ star produces \hbox{$\sim10^{37}~\mathrm{erg~s^{-1} \,
\mathrm{M}_\odot^{-1}}$} in the $\lya$ line, assuming $f_{\mathrm
esc}=0$.  At $z=7$ that would result in a flux of
\hbox{$\sim10^{-23}~\mathrm{erg~s^{-1}\,cm^{-2} \,
\mathrm{M}_\odot^{-1}}$}.  For a discussion of the size of the $\lya$
emitting region, see Loeb \& Rybicki \shortcite{loe99}.  For the
$f_{\mathrm esc}=1$ case the luminosity and size of the emitting
region of a \pop\ star depend in more detail on redshift and IGM
parameters.

Future theoretical work should improve our understanding of the
efficiency of \pop\ star formation, as well as the physical conditions
of the haloes in which \pop\ stars form.  We finally expect
the launch of {\it NGST} to open
an observational window into the earliest epochs of star
formation, possibly including \pop\ stars, less than a decade from now.

\section*{Acknowledgments}

VB thanks the TAPIR group at Caltech for its hospitality during the
completion of this work.  We thank L. Cambr\'{e}sy, W. T. Reach,
Y. Lithwick, A. Benson, J. Silk, A. Ferrara, M. Rees and P. Natarajan
for stimulating discussions.  We thank the anonymous referee for
comments that improved the presentation of this paper.  MRS
acknowledges the support of NASA GSRP grant NGT5-50339. This work was
supported in part at Caltech by NSF AST-0096023, NASA NAG5-8506 and
DoE DE-FG03-92-ER40701 and DE-FG03-88-ER40397.

\end{document}